\documentclass[aps,prl,twocolumn,groupedaddress]{revtex4-1}

\usepackage{graphicx}

\usepackage{epstopdf}
\usepackage{amsmath,amsfonts,amssymb}	

\usepackage{lineno}										

\setcitestyle{aysep={}}										
\newcommand{\field}[1]{\mathbb{#1}}   
\newcommand {\R}{\field{R} }							
\newcommand {\eps}{\varepsilon }					
\newcommand*\patchAmsMathEnvironmentForLineno[1]{%
  \expandafter\let\csname old#1\expandafter\endcsname\csname #1\endcsname
  \expandafter\let\csname oldend#1\expandafter\endcsname\csname end#1\endcsname
  \renewenvironment{#1}%
     {\linenomath\csname old#1\endcsname}%
     {\csname oldend#1\endcsname\endlinenomath}}%
\newcommand*\patchBothAmsMathEnvironmentsForLineno[1]{%
  \patchAmsMathEnvironmentForLineno{#1}%
  \patchAmsMathEnvironmentForLineno{#1*}}%
\AtBeginDocument{%
\patchBothAmsMathEnvironmentsForLineno{equation}%
\patchBothAmsMathEnvironmentsForLineno{align}%
\patchBothAmsMathEnvironmentsForLineno{flalign}%
\patchBothAmsMathEnvironmentsForLineno{alignat}%
\patchBothAmsMathEnvironmentsForLineno{gather}%
\patchBothAmsMathEnvironmentsForLineno{multline}%
}

\usepackage{color}
\usepackage{colordvi}

\newcommand{\komment}[1]{{}}

\begin{document}

\title{Fisher-Wright model with deterministic seed bank and selection
}


\author{Bendix Koopmann} 
\affiliation{Center for Mathematics, Technische Universit\"at M\"unchen, 85748 Garching, Germany}
\author{Johannes M\"uller} 
\affiliation{Center for Mathematics, Technische Universit\"at M\"unchen, 85748 Garching, Germany}
\affiliation{Institute for Computational Biology, Helmholtz Center Munich, 85764 Neuherberg, Germany}
\author{Aur\'elien Tellier} 
\affiliation{ Section of Population Genetics, Center of Life and Food Sciences Weihenstephan, Technische Universit\"at M\"unchen, 85354 Freising, Germany}
\author{Daniel \v Zivkovi\'c} 
\affiliation{ Section of Population Genetics, Center of Life and Food Sciences Weihenstephan, Technische Universit\"at M\"unchen, 85354 Freising, Germany}

%
%


\begin{abstract}
Seed banks are a common characteristics to many plant species, which allow storage of genetic diversity in the soil as dormant seeds for various periods of time. We investigate an above-ground population following a Fisher-Wright model with selection coupled with a deterministic seed bank assuming the length of the seed bank is kept constant and the number of seeds is large. To assess the combined impact of seed banks and selection on genetic diversity, we derive a general diffusion model. The applied techniques outline a path of approximating a stochastic delay differential equation by an appropriately rescaled stochastic differential equation, which is a common issue in statistical physics. We compute the equilibrium solution of the site-frequency spectrum and derive the times to fixation of an allele with and without selection. Finally, it is demonstrated that seed banks enhance the effect of selection onto the site-frequency spectrum while slowing down the time until the mutation-selection equilibrium is reached.  
\keywords{diffusion \and Fisher-Wright model \and seed bank \and selection \and site-frequency spectrum \and times to fixation}
\end{abstract}

\maketitle

\newpage
\section{Introduction}\label{intro}
Population genetics has intrinsic similarities with statistical physics~\cite{deVladarandBarton2011}, as aiming to describe the dynamics of two or several interacting types of individuals in a finite population. This formulation is intriguingly close to simple spin systems. Basic models in population genetics have been considered independently in statistical physics~\cite{Higgs95}. 
In particular, the Moran model has been investigated in the perspective of statistical physics~\cite{Aguiar11} e.g.\ with special attention to fluctuations~\cite{Lorenz2013} or fixation probabilities~\cite{Houchmandzadeh2010}. In the present work, we focus on the effect of delay in a population genetics context, and how to approximate such a model by an appropriate rescaled stochastic differential equation (SDE) without delay. \\
Stochastic delay differential equations (SDDEs) have wide-spread applications, e.g.\ in optics and laser physics, hydrodynamic processes, and various field of biological systems (see~\cite{Lafuerza2011} or in particular~\cite{Frank2005} and quotations therein).  
The derivation of SDDEs~\cite{Frank2007} and the appropriate  approximation of SDDEs by SDEs, e.g. for small delays, are discussed in~\cite{Guillouzic1999,Frank2016}. 
In the present article, we propose a method to cover delays in population genetics caused by seedbanks.\par\medskip

Dormancy of reproductive structures, that is seeds or eggs, is described as a bet-hedging strategy~\cite{evans2005,cohen1966}
in plants~\cite{honnay2008,evans2007a,tielboerger2012}, 
invertebrates, {\textit{e.g.}, Daphnia} \cite{decaestecker2007}, 
and microorganisms~\cite{lennon2011} 
to buffer against environmental variability. 
Bet-hedging is widely defined as an evolutionary stable strategy in which adults release their offspring into several different environments, here specifically with dormancy at different generations in time, to maximize the chance of survival and reproductive success, thus magnifying the evolutionary effect of good years and dampening the effect of bad years~\cite{evans2005,cohen1966}. 
Dormancy and quiescence sometimes have surprising and counterintuitive consequences, similar to diffusion in activator-inhibitor models~\cite{hadeler}. 
In the following study, we focus more specifically on the evolution of dormancy in plant species~\cite{honnay2008,evans2007a,tielboerger2012},   
but the theoretical models also apply to microorganisms and invertebrate species~\cite{decaestecker2007,lennon2011}.\par

Seed banking is a specific life-history characteristic of most plant species, which produce seeds remaining in the soil for short to long periods of time (up to several generations), and it has large but yet underappreciated consequences \cite{evans2005} for the evolution and conservation of many plant species.\par

First, polymorphism and genetic diversity are increased in a plant population with seed banks compared to the situation without banks. This is mostly due to storage of genetic diversity in the soil~\cite{kaj2001,nunney2002}. 
Seed banks also damp off the variation in population sizes over time~\cite{nunney2002}. 
Under unfavourable conditions at generation $t$, the small offspring production is compensated at the next generation $t+1$ by individuals from the bank germinating at a given rate. Under the assumption of large seed banks, the observed population sizes between consecutive generations ($t$ and $t+1$) may then be uncoupled. \par

Second, seed banks may counteract habitat fragmentation by buffering against the extinction of small and isolated populations, a phenomenon known as the ``temporal rescue effect''~\cite{brown1977}. 
Populations which suffer dramatically from events of decrease in population size can be rescued by seeds from the bank. Improving our understanding of the evolutionary conditions for the existence of long-term dormancy and its genetic underpinnings is thus important for the conservation of endangered plant species in habitats under destruction by human activities.\par

Third, germ banks influence the rate of natural selection in populations. On the one hand, seed banks promote the occurrence of balancing selection for example for color morphs in {\it Linanthus parryae} \cite{turelli2001} 
or in host-parasite coevolution~\cite{tellier2009}. 
On the other hand, the storage effect is expected to decrease the efficiency of positive selection in populations, thus natural selection, positive or negative, would be slowed down by the presence of long-term seed banks. Empirical evidence for this phenomenon has been shown~\cite{hairston1988}, 
but no quantitative model exists so far. In general terms, understanding how seed banks evolve, affect the speed of adaptive response to environmental changes, and determine the rate of population extinction in many plant species is of importance for conservation genetics under the current period of anthropologically driven climate change.\par\smallskip

Two classes of theoretical models have been developed for studying the influence of seed banks on genetic variability. First, \citet{kaj2001} have proposed a backward in time coalescent seed bank model which includes the probability of a seed to germinate after a number of years in the soil and a maximum amount of time that seeds can spend in the bank. 
Seed banks have the property to enhance the size of the coalescent tree of a sample of chromosomes from the above ground population by a quadratic factor of the average time that seeds spend in the bank. This leads to a rescaling of the Kingman coalescent~\cite{kingman1982} 
because two lineages can only coalesce in the above-ground population in a given ancestral plant. The consequence of longer seed banks with smaller values of the germination rate is thus to increase the effective size of populations and genetic diversity~\cite{kaj2001} 
and to reduce the differentiation among populations connected by migration~\cite{vitalis2004}. 
This rescaling effect on the coalescence of lineages in a population has also important consequences for the statistical inference of past demographic events~\cite{zivkovic2012}. 
In practice this means that the spatial structure of populations and seed bank effects on demography and selection are difficult to disentangle~\cite{boendel2015}. 
Nevertheless, \citet{tellier2011} could use this rescaled seed bank coalescent model~\cite{kaj2001} 
and Approximate Bayesian Computation to infer the germination rate in two wild tomato species {\it Solanum chilense} and {\it S. peruvianum} 
from polymorphism data~\cite{tellier2011a}. \par 
A second class of models assumes a strong seed bank effect, whereby the time seeds can spend in the bank is very long, that is longer than the population coalescent time~\cite{GonzalezCasanova201462}, 
or the time for two lineages to coalesce can be unbounded. This latest model generates a seed bank coalescent~\cite{blath2015a}, 
which may not come down from infinity and for which the expected site-frequency spectrum (SFS) may differ significantly from that of the Kingman coalescent~\cite{blath2015b}. 
In effect, the model of~\cite{kaj2001} 
represents a special case, also called a weak seed bank, where the time for lineages to coalesce is finite because the maximum time that seeds can spend in the bank is bounded.
\par

In the following we mainly have the weak seed bank model in mind where the time in the seed bank is bounded to a small finite number assumed to be realistic for most plant species~\cite{honnay2008,evans2007a,tielboerger2012,tellier2011a}. 
Even if we allow for unbounded times a seed may be stored within the soil, 
we assume that the germination probability decreases rapidly with age such 
that e.g.\ the expected time a seed rests in the soil is finite. 
We develop a forward in time diffusion for seed banks following a Fisher-Wright model with random genetic drift and selection acting on one of two genotypes. The time rescaling induced by the seed bank is shown to be equivalent for the Fisher-Wright and the Moran model. We provide the first theoretical estimates of the effect of seed bank on natural selection by deriving the expected SFS of alleles observed in a sample of chromosomes and the time to fixation of an allele.

The main difficulty in the present paper is the non-Markovian character of seedbank 
models (with the exception of a geometric survival distribution for seeds, in which case the 
model can be reduced to a Markovian model, see below). 
The way to deal with
this non-Markovian character is based on a separation of time scales. The genetic composition of the population only changes on a slow, so-called evolutionary time scale (thousands of generations), while being fairly stable on a fast, ecological time scale (tens of generations). We assume seeds to have a life span corresponding to this ecological time scale, and thus the seedbank tends to a quasi-stationary state. The non-Markovian character of the model is visible at the ecological time scale, while it vanishes on the evolutionary time-scale due to the quasi-steady-state assumption. In other words we ensure the separation of time scales by assuming that most seeds die after a few generations. We demonstrate thereafter that seed banks affect selection and genetic drift differently.
 
%


\section{Model description}

We consider a finite plant-population of  
size $N$. The plants appear in two genotypes $A$ and $a$. We assume non-overlapping generations. 
Let $X_n$ denote the number of type-$A$ 
plants in generation $n$ (that is, the number of 
living type-a plants in this generation is $N-X_n$). 
Plants produce seeds. The number of seeds is assumed to be large,
such that noise in the seed bank does not play a role (therefore we call the seed bank ``deterministic''). The amount
of seeds produced by type-$A$-plants in generation $n$ is $\beta_A X_n$, 
that of type-a plants $\beta_a (N-X_n)$. The seeds are stored \textit{e.g.} 
in the soil; some germinate in the next generation, some only in later generations, and some never.\par

To obtain the next generation of living plants $X_{n}$, we need to know which seeds are likely to germinate. Let $b_A(i)$ be the fraction of type-$A$ seeds of age $i$ able to germinate, and $b_a(i)$ that of type-a seeds. Hence, the total amount of type-$A$ seeds that is able to
germinate is given by
\[ 
\sum_{i=1}^\infty b_A(i)\beta_A X_{n-i},
\]
and accordingly, the total amount of all seeds that may germinate
\[ 
\sum_{i=1}^\infty b_A(i)\beta_A X_{n-i}
+
\sum_{i=1}^\infty b_a(i)\beta_a (N-X_{n-i}).
\]
The probability that a plant in generation $n$ is of phenotype $A$ is given by the fraction of 
type-A seeds that may germinate among all seeds that are able to germinate.
The frequency process of the di-allelic Fisher-Wright model with deterministic seed bank reads
\begin{eqnarray}
&& X_{n}\sim \mbox{Bin}(N, q_n(X_\bullet)),\label{WFseedModel}\\
&& q_n(X_\bullet) =\frac{\sum_{i=1}^\infty b_A(i)\beta_A X_{n-i}}
{\sum_{i=1}^\infty b_A(i)\beta_A X_{n-i}
+
\sum_{i=1}^\infty b_a(i)\beta_a (N-X_{n-i})}.\nonumber
\end{eqnarray}
Next we introduce (weak) selection.  The fertility of type-a plants is given by
\[ 
\beta_a =(1- s_1)\,\beta_A,
\]
such that $s_1=0$ corresponds to the neutral case. 
Furthermore, the fraction of surviving seeds is affected. We relate $b_a(i)$ to $b_A(i)$ by
\[ 
b_a(i) = (1-s_2)\,b_A(i).
\]
Of course, $s_2$ has to be small enough 
to ensure 
that $b_a(i)\in[0,1]$. There are other ways to 
incorporate a fitness difference in the surviving
probabilities of seeds, but we feel that this is
the most simple version. If we lump $s_1$ and $s_2$ in one parameter that scales in an appropriate way for selection,  
\[ 
(1-s_1)\,(1- s_2) = 1-\sigma/N,
\]
(the sign is chosen in such a way that genotype A has an advantage over genotype a for $\sigma>0$ and a disadvantage if $\sigma<0$) 
then eqn.\ (\ref{WFseedModel}) for $q_n(X_\bullet)$ with selection becomes
\[ 
  \frac{\sum_{i=1}^\infty b_A(i)X_{n-i}}
{\sum_{i=1}^\infty b_A(i) X_{n-i}
+ (1-\sigma/N)
\sum_{i=1}^\infty b_A(i) (N-X_{n-i})}.
\]
As this ratio is homogeneous of degree zero in $b_A$, we assume \mbox{$\sum_{i=1}^\infty b_A(i)=1$}. That is, $b_A(i)$ is considered a probability distribution for the survival of a (type-$A$) seed. We assume that the average life time of a seed is finite, 
$B=\sum_{i=1}^\infty i b_A(i)<\infty$. 
We will implicitly assume that $b_A(i)$ converge fast enough to zero, such that the 
separation of ecological and evolutionary time scale is still true. 
The sum $\sum_{i=1}^\infty b_A(i) X_{n-i}$ is a moving average. We emphasize this fact by introducing the operator
\[
M_n(X_\bullet) = \sum_{i=1}^\infty b_A(i) X_{n-i}.
\]
As a consequence, we have $M_n(N)=N$, and 
\begin{eqnarray}
 q_n(X_\bullet) &=& \frac{M_n(X_\bullet)}
{M_n(X_\bullet)
+ (1-\sigma/N) (N-M_n(X_\bullet))}\nonumber\\
 &=&\frac{M_n(X_\bullet)}
{
N-\sigma/N\, (N-M_n(X_\bullet))}.
\label{WFseedModel2}
\end{eqnarray}

\section{Diffusion limit -- geometric case}
As indicated above, if $b_A(i)$ follow a geometric distribution, then 
the non-Markovian model introduced above can be reduced to a Markovian model: it is not necessary to track the age of a seed, as all seeds independent 
of their age have the same mortality resp.\ germination probability. 
In this case, and without selection ($\sigma=0$), 
it is straight forward to obtain a diffusion limit that describes
the model well on the evolutionary time scale if the population size is (finite but) large. In particular, the diffusion limit is  
the diffusive Moran model, where we already obtain a first indication how the scaling is 
affected by the seedbank. Note that the backward process has been analyzed in~\cite{blath2013}. This neutral case with a geometric germination rate serves as a warm-up before investigating the full model.

\subsection{The Fisher-Wright model without selection}
We recall briefly the procedure to derive the diffusion limit 
for the standard Fisher-Wright model (without seed bank).\\
$\bullet$ {\it Model:} $ X_{n+1} \sim \mbox{Bin}(N, X_n/N)$.\\
$\bullet$ {\it Rescale population size:} 
Let $x_n =X_n/N$. Then, 
$ X_{n+1} \sim \mbox{Bin}(N, x_n).$ For $N$ large, the Binomial distribution 
approximates a normal distribution with expectation $x_n\, N$ and variance $x_n(1-x_n) N$. 
Let $\eta_n$ be i.i.d. $N(0,1)$-random variables. Then,
\begin{eqnarray*}
x_{n+1} = X_{n+1}/N &\approx{}& \left( x_n\, N + (x_n(1-x_n))^{1/2} N^{1/2} \eta_n\right)/N\\
 &=& x_n  + N^{-1/2}\,\,(x_n(1-x_n))^{1/2} \,\, \eta_n.
\end{eqnarray*}
$\bullet$ {\it Rescale time:} 
Now  define $\Delta \tau = 1/N$, introduce the time $\tau=n\Delta \tau$, let $u_{n\Delta \tau} = x_n$,  and rescale the index of the normal random variables, that is, replace $\eta_n$ by $\eta_{n\Delta {\tau}}=\eta_\tau$. Then, 
$ u_{\tau+\Delta \tau}-u_\tau  =   \Delta \tau^{1/2}\,\,(u_\tau(1-u_\tau))^{1/2} \,\, \eta_\tau$.
According to the Euler-Maruyama formula (see \textit{e.g.}~\cite{kloeden:sde}),  we approximate the 
diffusive Moran model for $N$ large (that is, $\Delta \tau=1/N$ small)
\[ 
du_\tau = (u_\tau(1-u_\tau))^{1/2} \, dW_\tau.
\] 
where $W_t$ indicates the Brownian motion.

\subsection{Seed bank model with a geometric germination rate and without selection}

In the present section we assume that there is no selection ($\sigma=0$), and 
$b(i)$ follow a geometric distribution with parameter \mbox{$\mu\in (0,1)$}, $b(1)=\mu$ and $b(i)=$\mbox{$(1-\mu)$}\mbox{$b(i-1)$}. In this case, 
the delay-model is equivalent to a proper Markov chain.\\
$\bullet$ {\it Reformulation of the model:} 
 Define $z_n=M_{n+1}(X_\bullet)/N=\mu\sum_{i=1}^\infty{}(1-\mu)^{i-1}X_{n+1-i}/N$. We immediately obtain
\begin{eqnarray*}
z_{n+1} &=& \mu\,
 \sum_{i=1}^\infty (1-\mu)^{i-1} X_{n+2-i}/N\\
 &=& \mu\,X_{n+1}/N + \mu\,\sum_{i=2}^\infty (1-\mu)^{i-1} X_{n+1-(i-1)}/N\\
 &=&  \mu\,X_{n+1}/N + (1-\mu)\,z_n.
\end{eqnarray*}
 Next (and with the nomenclature of (\ref{WFseedModel2})), we have
$q_{n+1}(X_\bullet)  = M_{n+1}(X_\bullet/N) = z_n$.
All in all, we reformulated  model (\ref{WFseedModel}) in the present situation as 
\begin{eqnarray}
X_{n+1} &\sim&\mbox{Bin}(N, z_n),\\
z_{n+1} &=&  \mu\,X_{n+1}/N + (1-\mu)\,z_n.\nonumber
\end{eqnarray}
Note that $z_n$ can be interpreted as the state of the seed bank 
(the fraction of type-$A$ seeds that are able to germinate).\par

$\bullet$ {\it Rescale population size:} As this model is Markovian, it is simple to 
derive the diffusion limit. As usual, we start off by defining $x_n = X_n/N$, and obtain 
$z_{n} = \mu\,x_n+(1-\mu)\, z_{n-1}$, $X_{n+1} = \mbox{Bin}(N, z_n)$. 
Approximating the Binomial distribution by a normal distribution for $N$ large yields
\[
x_{n+1} \approx  z_n + N^{-1/2} (z_n(1-z_n))^{1/2} \eta_n,
\]
where the $\eta_n\sim N(0,1)$ i.i.d.. As $x_{n+1}$ can be expressed by $z_n$ and $z_{n+1}$, 
the foregoing two equations give 
\[
\frac {z_{n+1} - (1-\mu) \, z_n}{\mu } = z_n  + N^{-1/2} (z_n(1-z_n))^{1/2} \eta_n.
\]
Therefore, $z_{n+1}-z_n = \mu \,N^{-1/2} \,(z_n(1-z_n))^{1/2}\,\eta_n$.\\
$\bullet$ {\it Rescale time:}  Scaling time by $N$ yields for $u_{n/N} = z_n$ and $\tau=n/N$
\[
du_\tau = \mu\, (u_\tau\,(1-u_{\tau}))^{1/2} dW_\tau.
\] 
If we define $B=1/\mu$ (the expected value of a geometric 
distribution with parameter $\mu$), we may write this equation as
\begin{eqnarray}
du_\tau =  \frac{(u_\tau\,(1-u_\tau))^{1/2}}{B} \,\, dW_\tau.
\end{eqnarray}
We find a diffusive Moran model for the state of the 
seed bank 
with rescaled time scale. The factor $1/B$ has been already proposed in 
the paper of Kaj, Krone and Lascoux~\cite{kaj2001}, who analyzed a 
seedbank process backward in time.

\section{Diffusion limit -- general case}

We expect a similar result as above to hold 
in the general case. A difference between the two cases is that we
naturally considered the state of the seed bank before, while in the
general case we will focus on the state of living plants. As discussed
before, the center of the analysis below is an additional step that 
investigates the quasi-stationary state of the seedbank at evolutionary time scale; this additional step is necessary to deal with the non-Markovian character of our model.

\subsection{Rescale population size}
From (2), we immediately have 
\[
q_{n}(x_\bullet) = q_{n}(X_\bullet/N) = \frac{M_n(x_\bullet)}{1-\Delta t\,\sigma(1-M_n(x_\bullet))}.
\]
Using Normal approximation of the Binomial distribution leads to 
\[
 x_n  \approx  q_n(x_\bullet)+\Delta t^{1/2}\sqrt{q_n(x_\bullet)\,(1-q_n(x_\bullet))}\, 
 \eta_n
\]
where $\eta_n\sim N(0,1)$ are i.i.d.. Taylor expansion of $q_n(x_\bullet)$  w.r.t.\ $\Delta t$ yields 
in lowest order 
\begin{eqnarray}
&& x_n-M_n(x_\bullet) -\Delta t\,\sigma\,f(M_n(x_\bullet))
\label{recEquXXX}\\
  &=&   \Delta t^{1/2}  f^{1/2}(M_n(x_\bullet))\, \eta_n. 
   \nonumber
 \end{eqnarray}
    with $f(x)=x(1-x)$. 

\subsection{Perturbation approach}
The leading term of eqn.\ (\ref{recEquXXX})  is
$x_n-M_n(x_\bullet)$. 
This difference must not become too large, as all other terms in the equation are at least of order $\Delta t^{1/2}$. That is, the state $x_n$ 
can only slowly drift away from 
$M_n(x_\bullet)$ (which represents the state of the seed bank). Hence, for a reasonable number of time steps (on the ecological time scale), $M_n(x_\bullet)$ is fairly constant. In order to 
disentangle the evolutionary and the ecological time scale, 
we introduce $\eps = \Delta t^{1/2}$, expand $x_n$ w.r.t. $\eps$,
$$ x_n = x_n^{(0)} + \eps x_n^{(1)}+ \eps^2 x_n^{(2)} + \ldots$$
and rewrite eqn. (\ref{recEquXXX}) as 
$$ 
 x_n-M_n(x_\bullet) = \eps^2 \,\sigma\,f(M_n(x_\bullet))
 +   \eps\, f^{1/2}(M_n(x_\bullet))  \eta_n. 
   $$
Taylor expansion and equating equal powers of $\eps$ yields
\begin{eqnarray}
&&x_n^{(0)} - M_n(x^{(0)}_\bullet) = 0\label{order0}\\
&&x_n^{(1)} - M_n(x^{(1)}_\bullet) =  f^{1/2}(M_n(x_\bullet^{(0)}))  \eta_n\label{order1}\\
&&x_n^{(2)} - M_n(x^{(2)}_\bullet) = \sigma(f(M_n(x_\bullet^{(0)}))\label{order2}\\
&&\qquad\qquad\quad +\frac 1 2 f^{-1/2}(M_n(x_\bullet^{(0)}))\, f'(M_n(x_\bullet^{(0)}))\, M_n(x_\bullet^{(1)})\, \eta_n.\nonumber
\end{eqnarray}
{\bf Zero order:} The zero order term~$x^{(0)}_n$ follows a deterministic dynamics. As 
$M_n$ is an averaging operator the solution becomes constant in the long run. The system saddles on 
the slow manifold, consisting of constant sequences. 
At this point it is important that $b_A(i)$ tend 
fast enough to zero, s.t.\ $x_n^{(0)}$ indeed approximates on the fast (ecological) time scale  the slow manifold. 
We assume $x_n^{(0)}\equiv \overline x^{0}$. \\
{\bf First order:}  
 The recursive equation (\ref{order1}) is well known as an auto-regression (AR) model in the statistical modeling of time series~\cite{brockwelldavis}. 
We define $\beta = f^{1/2}(M_n(x_\bullet^{(0)})) =f^{1/2}(\overline x^0)$ (note that $\beta$ is a real number and not a random variable) and 
convert the AR model into a moving average equation. 
Thereto we introduce the back-shift operator acting on the index of a sequence, 
 $L z_n = z_{n-1}$, 
and the power series  
\[
\psi(x) = 1-\sum_{i=1}^\infty b_A(i)x^i;
\]
Eqn. (\ref{order1}) becomes in this notation
\[
\psi(L) x^{(1)}_{n} = x^{(1)}_{n}-M_n(z_\bullet) =  \,\beta\,  \eta_n.
\]
Note that $\psi(1)=0$, which does mean that the AR model is non-stationary (this process is also called an ARIMA model for time series~\cite[Chapter 9]{brockwelldavis}). We do not find 
a power series $\psi^\ast(x)$ well defined at $x=1$ such that $\psi^\ast(x)\,\psi(x)=1$. Therefore, 
we rewrite $\psi(x)$ as 
$\psi(x) = (1-x)\,\tilde\psi(x)$ (which is the defining equation of $\tilde\psi(x)$). As 
\[
\tilde \psi(1) =  \lim_{x \rightarrow 1}\frac{\psi(x)}{(1-x)}
=-\psi'(1)= \sum_{i=1}^\infty b_A(i)\, i = B\not=0,
\] 
we do find $\psi^\ast(x)$ such that $\psi^\ast(x)\tilde\psi(x)=1$, 
and hence $\psi^\ast(x)\psi(x)=1-x$ in a neighbourhood of $x=1$. 
As an immediate consequence (used later) we have 
 $\psi^*(1)=1/B$. 
If we multiply the equation $\psi(L) x^{(1)}_{n} = \,\beta\,  \eta_n$ 
by $\psi^*(L)$, we obtain 
\[
x^{(1)}_{n}-x^{(1)}_{n-1} = (1-L)x^{(1)}_{n} 
= \beta \psi^\ast(L)\eta_n
\]
and 
\begin{eqnarray*}
x^{(1)}_{n} & = &x^{(1)}_{n-1} +  \,\beta\,\psi^*(L) \eta_n\\
&=& x^{(1)}_{n-2} +  \,\beta\,\psi^*(L) \eta_n+  \Delta t^{1/2} \,\beta\,\psi^*(L) \eta_{n-1} =\cdots \\
&\approx{}&   \,\beta\, \sum_{\ell=0}^n \psi^*(L)\eta_{n-\ell}.
\end{eqnarray*}
Let  $\psi^*(z) = \sum_{i=0}^\infty a_i z^i$. 
We expand the sum above, and obtain
\begin{widetext}
\begin{eqnarray*}
\begin{array}{cccclllllll}
\sum_{\ell=0}^n \psi^*(L)\eta_{n-\ell} & = & a_0\eta_{n} &+ a_1 \eta_{n-1} &+ a_2 \eta_{n-2} &+ a_3 \eta_{n-3} &+ a_4 \eta_{n-4} &+ a_5 \eta_{n-5} &+\cdots\\
  &     &                 &+ a_0 \eta_{n-1} &+ a_1 \eta_{n-2} &+ a_2 \eta_{n-3} &+ a_3 \eta_{n-4} &+ a_4 \eta_{n-5}   &+\cdots\\
  &     &                 &                 &+ a_0 \eta_{n-2} &+ a_1 \eta_{n-3} &+ a_2 \eta_{n-4} &+ a_3 \eta_{n-5}  & +\cdots\\
  &     &                 &                 &                 &+ a_0 \eta_{n-3} &+ a_1 \eta_{n-4} &+ a_2 \eta_{n-5} & +\cdots\\\
  &     &                 &                 &                 &                 &+\cdots&+\cdots&+\cdots
\end{array}
\end{eqnarray*}
\end{widetext}
If we inspect not rows (that have $\psi^*(L)\eta_{i-\ell}$ as entries) but columns (that contain
always the same random variable $\eta_{i-\ell}$), we find that the coefficient
in front of one given random variable $\eta_{i-\ell}$ approximates $\psi^*(1)$ for $\ell\rightarrow\infty$.
\par
At this point, we want to write 
$ x^{(1)}_{n+1} \approx  \,\beta\, \,\psi^*(1) \sum_{\ell=1}^n \eta_{\ell}$.
This is only true, also in an approximate sense, 
 if $n$ is large and the state $x_n$ does hardly change over a time scale that allows 
$\sum_{i=1}^m a_i$ to converge to $\psi^*(1)=1/B$. 
If $\Delta t^{1/2}$ is small, then indeed 
$x_n \approx \overline x_0$ on the ecological time scale, as required. 
Hence, for $\Delta t$ small we are allowed to 
assume
\[ 
x^{(1)}_{n+1} \approx \,\beta\, \,\psi^*(1) \sum_{\ell=1}^n \eta_{\ell}
=  \frac{\beta}{B}\,\, \sum_{\ell=1}^n \eta_{\ell}.
\]
Thus, $ x^{(1)}_{n+1} \approx (\beta/B) \, \sum_{\ell=1}^n \eta_{\ell}$, and 
 for $n$ large 
\begin{eqnarray}
x^{(1)}_{n+1}-x^{(1)}_n = (\beta/B) \,  \eta_{n}\label{resulstOrder1}
\end{eqnarray}
where, as before, $\eta_n\sim N(0,1)$ i.i.d..

{\bf Second order:} 
With $\alpha=f(\overline x^0)$, 
$\tilde \beta=\frac 1 2 f^{-1/2}(\overline x^0)\, f'(\overline x^0))$,  we may write
\begin{eqnarray}
x^{(2)}_{n}+M_n(x^{(2)}_\bullet)=\alpha +\tilde \beta\,M_n(x^{(1)}_\bullet) \eta_{n-1}.\label{xxxCorXX}
\end{eqnarray}
If $\alpha\not = 0$, $x^{(2)}_n$ incorporates a 
deterministic trend. We first remove this trend defining 
 $z_n = x^{(2)}_n-w_n$ with $w_n = n\,\alpha/B$.
Then, $M_n(w_\bullet) = \sum_{i=1}^\infty b_A(i)\,(n-i)\,\alpha/B
= n\,\alpha/B - \alpha$, and, 
with $M_n(x^{(2)}_\bullet) = M_n(z_\bullet) + M_n(w_\bullet)$,
\begin{eqnarray*}
&&z_n + n\,\frac\alpha B -\bigg( M_n(z_\bullet) + M_n(w_\bullet) \bigg) -\alpha\\
&&=   z_n -M_n(z_\bullet) + n\,\frac\alpha B -\alpha - \left( n\,\frac \alpha  B - \alpha\right) 
=  \tilde \beta\, M_n(x^{(1)}_\bullet) \eta_n.
\end{eqnarray*}
We obtain an AR model for $z_n$ without trend, 
\begin{eqnarray}
z_{n}-M_n(z_\bullet) =  
 \,\tilde \beta\, M_n(x^{(1)}_\bullet) \eta_n.
\end{eqnarray}
It turns out, that we need not to analyze $z_n$ in detail. It is sufficient to note that $z_n$ is a random variable with expectation zero.\par\medskip
{\bf Result: }All in all, we conclude
$$ x_{n+1}-x_n = \eps^2\,\frac \alpha B + \eps\frac{ \beta} B\, \eta_n + \eps^2 z_n.$$
We only  take into account the lowest order in 
the deterministic drift resp.\ in the random perturbations. As $\eps\,(\beta/B)\,\eta_n$ dominates 
$\eps^2\,z_n$,  we drop the latter term, replace 
in $\alpha$, $\beta$ the variable $\overline x^0$ by $x_n$, and
end up with  
\begin{eqnarray}
x_{n} &=& x_{n-1} + \Delta t \frac \sigma B x_n(1-x_n)\\
&& \qquad\quad+ \Delta t^{1/2} \frac 1 B \,\, \sqrt{\,x_n(1-x_n)}\,\eta_n.\nonumber
\end{eqnarray}
\begin{figure*}[htbp]
\begin{center}
\includegraphics*[width=0.46\textwidth]{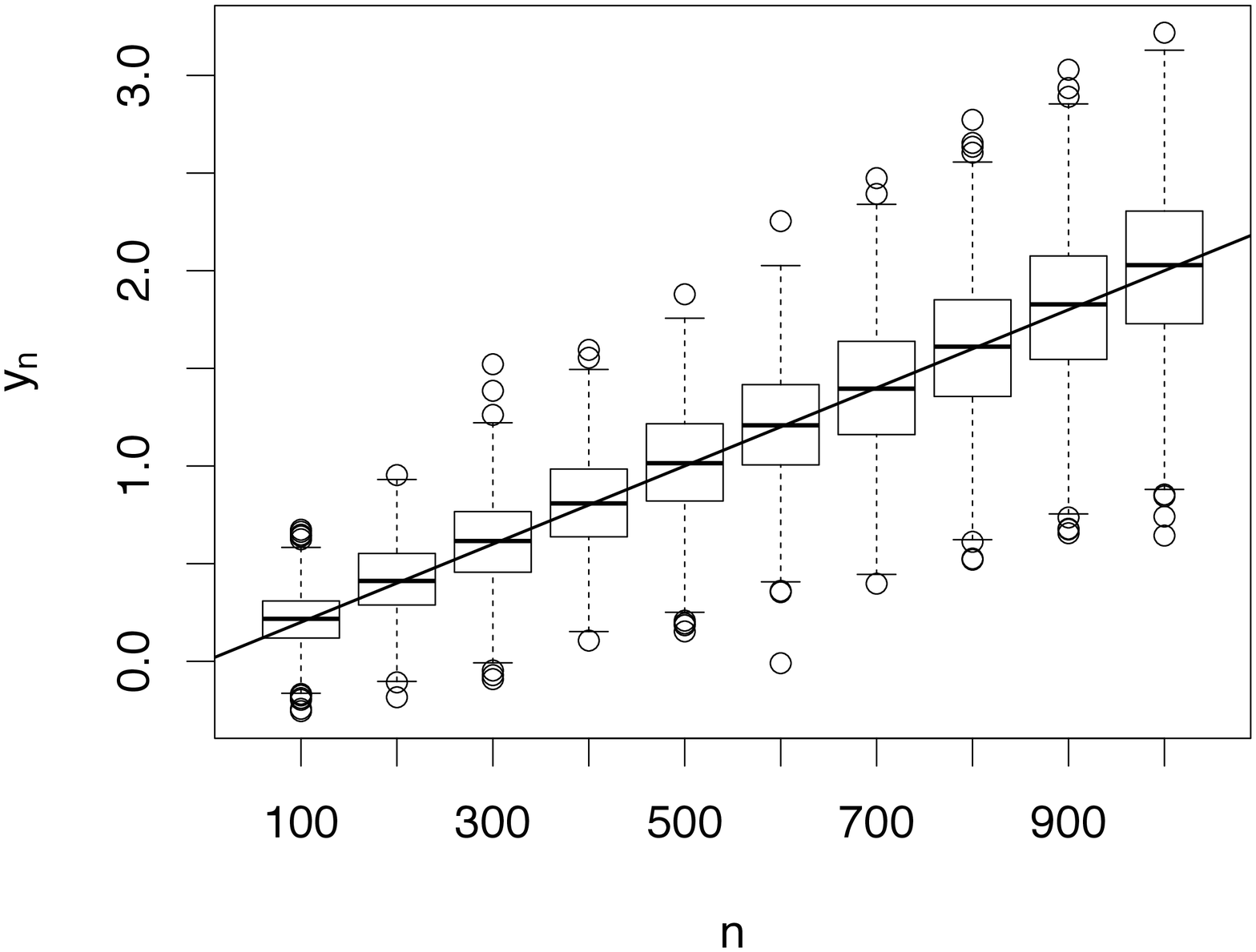}\hspace*{1cm}
\includegraphics*[width=0.46\textwidth]{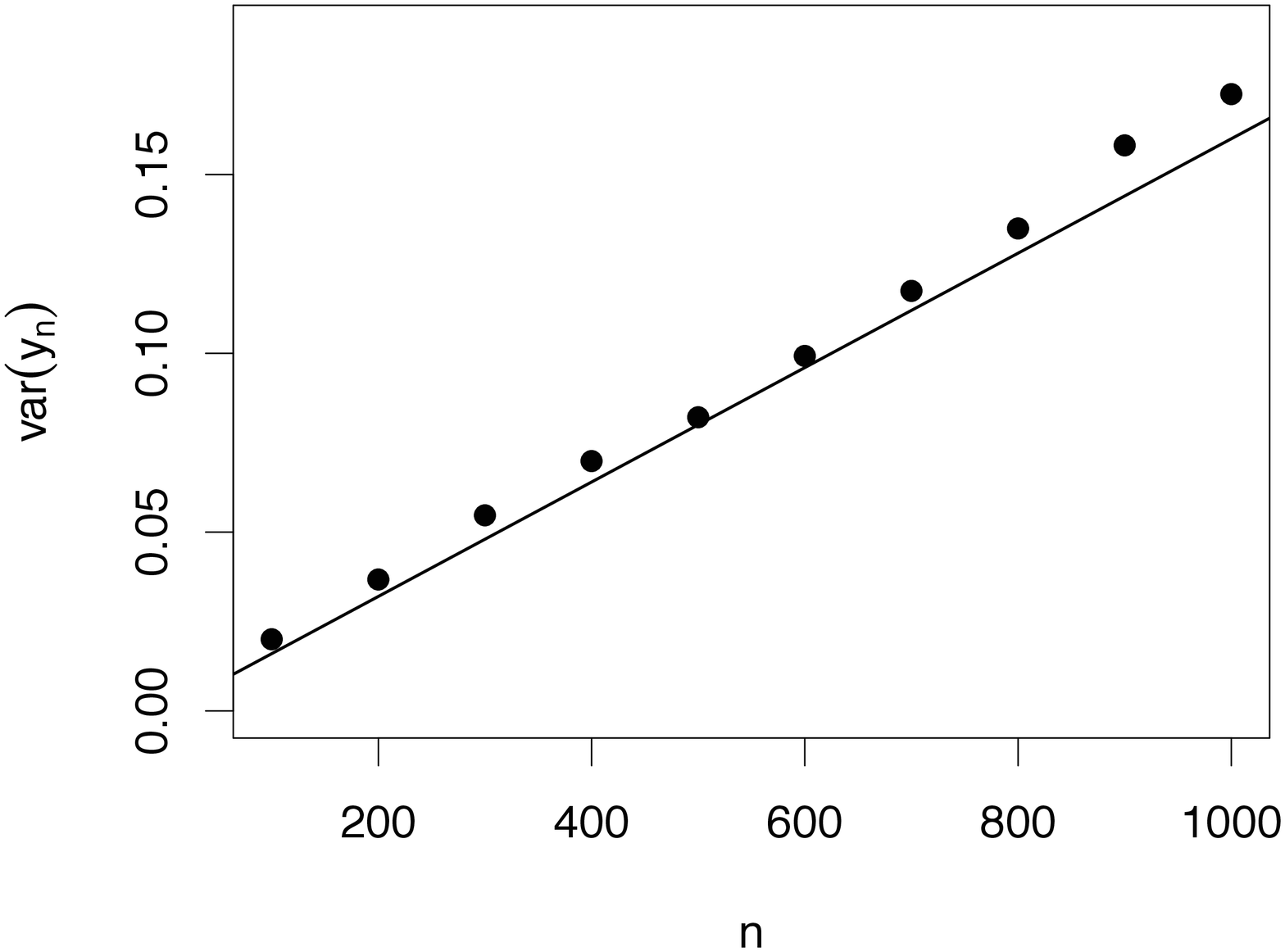}
\end{center}
\caption{Simulation of the AR model (1000 runs). Samples have been taken at time steps $100$, $200$,$\ldots$, 1000. (left) Boxplot of the simulated time series $y_n$ at indicated time points together with 
the mean according to eqn.~\ref{xxxCorXX} (line). (right) 
Variance of the simulated time series at indicated time points (dots), together with the variance  according to eqn.~\ref{xxxCorXX} (line).  For parameters used: see text.}\label{simMAfig}
\end{figure*}

{\bf Numerical simulation:} We compare the result of these computations with numerical simulations. 
Thereto we consider the linear model
$$ y_n - M_n(y_\bullet) = \Delta t\,a\,+\,\Delta t^{1/2}\,b\,\eta_n$$
with $a, b\in \R$. 
If $y_n=0$ for $n\leq 0$, we expect that $y_n$ 
(for $n\geq 1$) approximately to satisfy 
$$ y_n - y_{n-1} = \Delta t\frac{a} B\,+\,
\Delta t^{1/2} 
 \,\frac{b}{B}\,\eta_n.$$
That is, $y_n$ is 
approximately normally
distributed with expectation $n\,\Delta t\,a/B$, and variance 
$n\,\Delta t\,b^2/B^2$. For simulations, we choose $a=1$, $\Delta t=0.01$, $b=2$ and $M_n(y_\bullet) = \frac 1 {m} \sum_{i=1}^{m} y_{n-i}$ for $m=9$, that is, $B=5$.  The 
simulations show an excellent agreement with our computations (Figure~\ref{simMAfig}).\\

\subsection{Rescale time}
As before, we define $u_{n\,\Delta t} = x_n$, and 
use the Euler-Maruyama-formula to conclude that $u_t$ 
approximates for $\Delta t\rightarrow 0$ the stochastic
differential equation 
\begin{eqnarray}
 du_t = 
 \frac \sigma B \,u_t(1-u_t) dt +\frac 1 B\,
 \bigg(u_t(1-u_t)\bigg)^{1/2}\, dW_t. \label{moranDiffProcess}
\end{eqnarray}
Please note that this result seems to inherit the
usual stability of a diffusion limit 
w.r.t.\ the detailed model assumptions: 
if we start off with a Moran model instead of a Fisher-Wright model combined with a seed bank, we again obtain a diffusion
limit of similar form (see Appendix).

We now change the time scale such that the variance coincides with the standard diffusive Moran model.
If we define $\tau = t/B^2$, then the SDE reads
\begin{eqnarray}
 du_\tau = 
 (\sigma B) \,u_\tau(1-u_\tau) d\tau +
 \sqrt{u_\tau(1-u_\tau)}\, dW_\tau.\label{rescaledDiffProcess}
\end{eqnarray}

{\bf Scaling of the selection parameter.} 
We conclude, in line with previous findings (see discussion), that the appropriate scaling of time 
for the Fisher-Wright model with seed bank is not 
$1/N$ but $1/(B^2\,N)$. Moreover, the effective selection rate 
(w.r.t.\ this time) is 
increased by the average number of generations $B$ 
the seeds sleep in the soil. 
\par

\section{The forward diffusion equation for seed bank models with selection}

In analogy to above, we consider a single locus and two allelic types $A$ and $a$ with frequencies $x$ and $1-x$, respectively, at time zero. Time is scaled in units of $2N$ generations. In the diffusion limit, as $N\to\infty$, the probability $f(y,t)dy$ that the type-A genotype has a frequency in $(y,y+dy)$ is characterized by the following forward equation (see \cite{Kimura1955} for $B=1$):
\begin{eqnarray*}
\frac {\partial}{\partial t} f(y,t) 
&=& 
- \frac {\partial}{\partial y}\left(
a(y)\,f(y,t)\right)
+\frac 1 2  \frac {\partial^2}{\partial y^2}
\left(
b(y)\,
f(y,t)\right),
\end{eqnarray*}
where the drift and the diffusion terms are given by $a(y)=\sigma\;y(1-y)/B$ and $b(y)=y(1-y)/B^2$, respectively. \par

For the derivations of the frequency spectrum and the times to fixation we require the following definitions. The scale density of the diffusion process is given by 
\[
\xi(y)=\exp\left(-\int_{0}^{y}\frac{2a(z)}{b(z)}dz\right)=\exp\left(-2 B \sigma y\right).
\]
The speed density is obtained (up to a constant) as
\[
\pi(y)=[b(y)\xi(y)]^{-1}=\frac{B^2 \exp\left(2 B \sigma  y\right)}{y (1-y)}.
\]
The probability of absorption at $y=0$ is given by 
\[
u_0(x)=\frac{\int_x^1\xi(z)dz}{\int_0^1\xi(z)dz}=\frac{\exp(2 B \sigma (1 - x))-1}{\exp(2 B \sigma)-1}, 
\]
and $u_1(x)=1-u_0(x)$ gives the probability of absorption at $y=1$. 

\subsection{Site-frequency spectra}

The site-frequency spectrum (SFS) of a sample (\textit{e.g.}, \cite{Griffiths2003,zivkovic2011,etheridge}) is widely used for population genetics data analysis. 
 A sample of size $k$ is sequenced, and for each polymorphic site the number of individuals in which the mutation appears is determined. In this way, a dataset is generated that summarizes the number of mutations $\zeta_{k,i}$ appearing in $i$ individuals, $i=1,\ldots,k-1$. That is, $\zeta_{k,1}=10$ indicates that $10$ mutations only appeared once, and $\zeta_{k,2}=5$ tells us that five mutations were present in two individuals (where the pair of individuals may be different for each of the five mutations). Note that neither $\zeta_{k,0}$ nor $\zeta_{k,k}$ are sensible: a mutation that appears in none or all individuals of the sample cannot be recognized as a mutation. In practice, it is often not possible to know the ancestral state. Then the folded SFS $\eta_{k,i}=(\zeta_{k,i}+\zeta_{k,k-i})(1+1_{\{i=k-i\}})^{-1}$ can be used. 
Since both empirical observations and theoretical results for the folded SFS follow instantaneously from the unfolded one, we only consider the unfolded version.\par

For the derivation of the theoretical SFS, we assume that mutations occur according to the infinitely-many sites model \cite{Kimura1969}. The scaled mutation rate is given by $\theta=4N\,\nu$, where $\nu$ is the mutation rate per generation at independent sites. Assuming that each mutant allele marginally follows the diffusion model specified above, the proportion of sites where the mutant frequency is in $(y,y+dy)$ is given by \cite{Griffiths2003}
\begin{eqnarray*}
\hat{f}(y)=\theta\,\pi(y)\,u_0(y)&=&\frac{\theta B^2}{y(1-y)}\,\frac{\exp(2 B \sigma)-\exp(2 B \sigma y)}{\exp(2 B \sigma)-1}\\
&=&\frac{\theta B^2}{y(1-y)}\,\frac{1-\exp(-2 B \sigma (1-y))}{1-\exp(-2 B \sigma)},
\end{eqnarray*}
where $\hat{f}(y)$ denotes the equilibrium solution of the population SFS. For neutrality, we immediately obtain $\hat{f}(y)=\theta\,B^2/y$ by letting $\sigma\to{}0$ in the foregoing equation.

The equilibrium solution of the SFS for a sample of size $k$ is obtained via binomial sampling (see \cite{zivkovic2015} for $B=1$) as 
\begin{eqnarray*}
\label{SFS}
\hat{f}_{k,i} &=& {k \choose  i} \, \int_0^1 \hat{f}(y) y^i(1-y)^{k-i}\, dy\\
&=&\theta{}B^2\frac{k}{i(k-i)}\frac{1-{}_1F_1(i;k;2B\sigma)e^{-2B\sigma}}{1-e^{-2B\sigma}},
\end{eqnarray*}
where $_1F_1$ denotes the confluent hypergeometric function of the first kind \cite{abramowitzandstegun}. For neutrality, we again immediately obtain $\hat{f}_{k,i}=\theta\,B^2/i$ by letting $\sigma\to{}0$. For a large number of mutant sites, the relative SFS $\hat{r}_{k,i}=\hat{f}_{k,i}/\sum_{j=1}^{k-1}\hat{f}_{k,j}$ approximates the empirical distribution $\zeta_{k,i}/\sum_{j=1}^{k-1}\zeta_{k,j}$ for a constant population size. Note that the solutions for the absolute SFS assume that mutations can occur at any time. When assuming that mutations can only arise in living plants \cite{kaj2001}, $\theta$ has to be replaced by $\theta/B$ in the respective equations. Both mutation models give equivalent results for the relative SFS.\par

\begin{figure*}[htbp]
\begin{center}
\includegraphics*[width=0.45\textwidth]{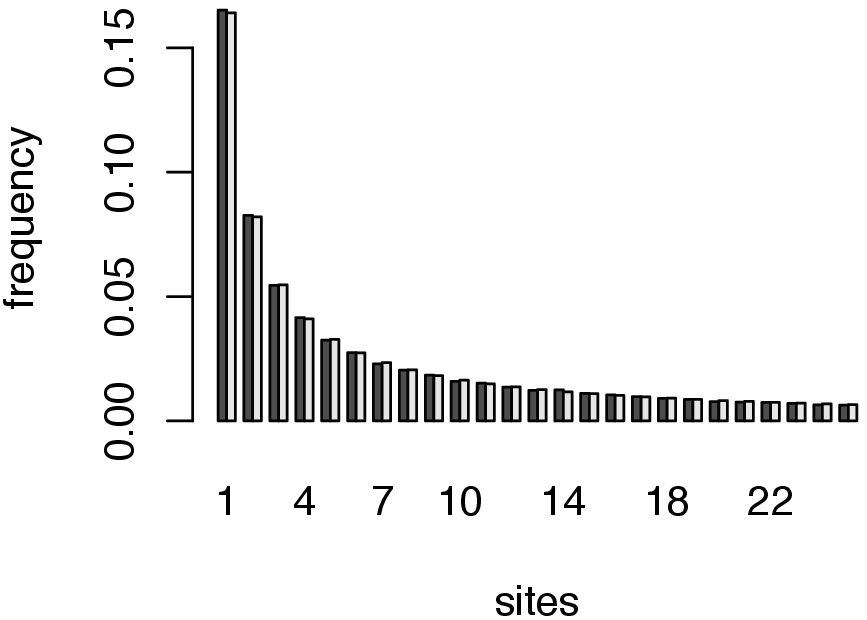}\hspace*{1cm}
\includegraphics*[width=0.45\textwidth]{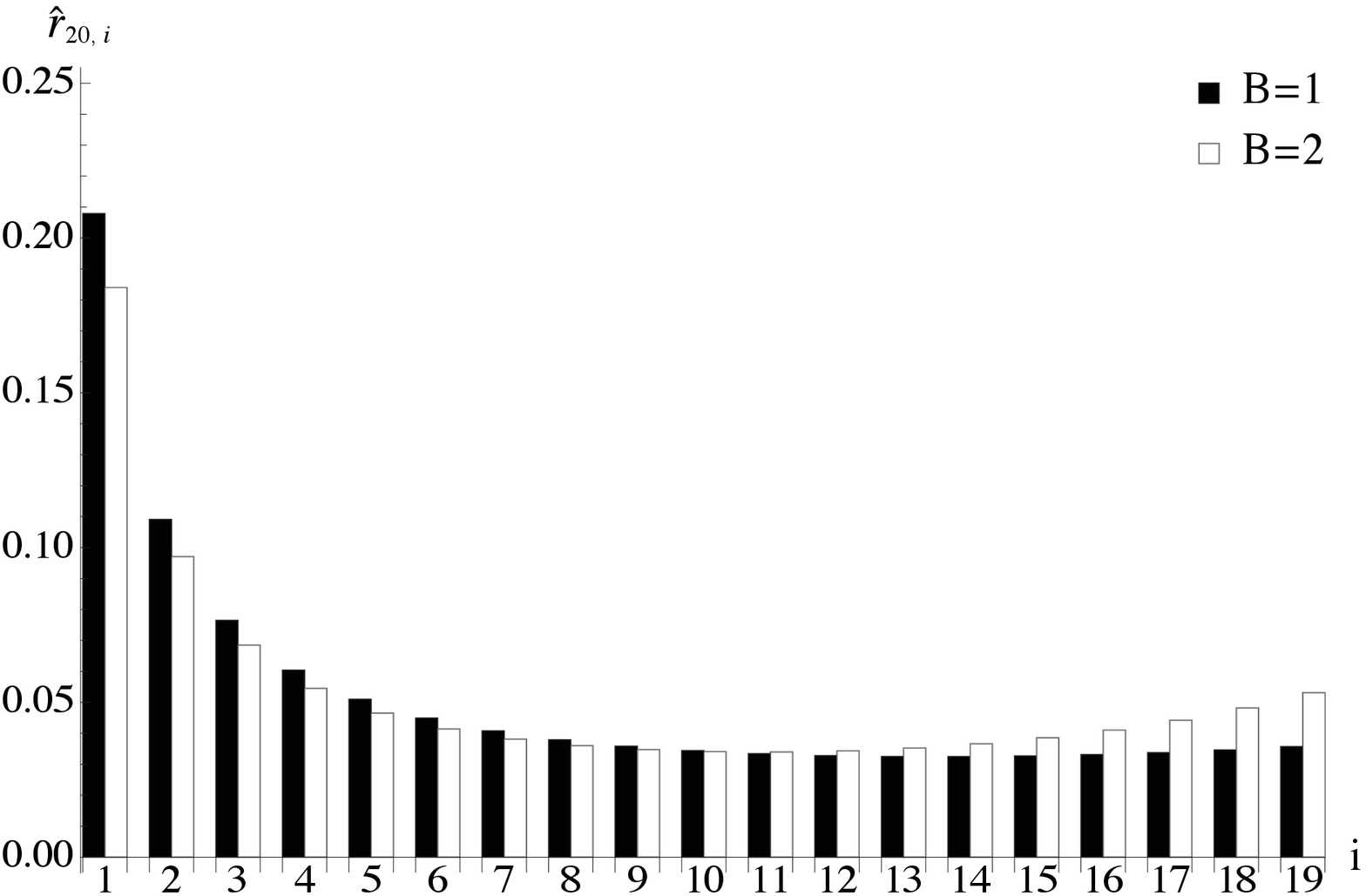}
\end{center}
\caption{(left) Simulation and theoretical prediction for the neutral relative SFS and a uniformly distributed seed bank of length $B=10$. For the simulation of the original discrete model the population size was chosen as 1000, we started without mutations and stopped the process after 400,000 generations to calculate the SFS as an average over 10,093 repetitions. The light gray bar shows the theoretical result, the dark gray bar shows the simulation outcome. In both cases a sample of 250 individuals was drawn. (right) Theoretical results for the relative SFS of a sample of size 20 are plotted for positive selection of strength $\sigma=2$ without ($B=1$) and with a seed bank of length $B=2$.
}\label{simSeedSfsSimVsTheo}
\end{figure*}

As shown in Figure~\ref{simSeedSfsSimVsTheo} (left), the neutral diffusion approximation is in line with the simulation results of the original discrete model. The theoretical relative SFS for a sample of 250 individuals approximates the simulated SFS, which is obtained as an average over 10,093 repetitions. In every iteration, the sample is drawn from an initially monomorphic population of 1000 individuals after 400,000 generations (so that the population has reached an equilibrium). Figure~\ref{simSeedSfsSimVsTheo} (right) illustrates the enhanced effect of selection proportional to the length of the seed bank.  

\subsection{Times to fixation}

We assume that both $y=0$ and $y=1$ are absorbing states and start by considering the mean time until one of these states is reached in the diffusion process specified above. The mean absorption time $\bar{t}$  can be expressed as~\cite{ewens}
\begin{equation}
\label{fixtime}
\bar{t}(x)=\int\limits_0^1{}t(x,y)dy, 
\end{equation}
where
\begin{eqnarray*}
t(x,y)&=&2\,u_0(x)[b(y)\xi(y)]^{-1}\int\limits_0^y{\xi(z)dz}, \quad{}0\leq{}y\leq{}x,\\
t(x,y)&=&2\,u_1(x)[b(y)\xi(y)]^{-1}\int\limits_y^1{\xi(z)dz}, \quad{}x\leq{}y\leq{}1.\\
\end{eqnarray*}
For genetic selection the integral in~(\ref{fixtime}) cannot be analytically solved. For selective neutrality, we obtain $\bar{t}(x)=-2\,B^2\,(x\,\log(x)+(1-x)\,\log(1-x))$ (see \textit{e.g.}~\cite{ewens} for $B=1$) by employing the drift term, the scale density and the probabilities of absorption as specified above.\par

Now, we evaluate the time until a mutant allele is fixed conditional on fixation as $\bar{t^*}(x)=\int_0^1{}t^*(x,y)dy$, where $t^*(x,y)=t(x,y)u_1(y)/u_1(x)$. For genic selection the mean time to fixation in dependency of x can only be derived as a very lengthy expression in terms of exponential integral functions. The neutral result is found as $\bar{t^*}(x)=-2 B^2(1 - x)/x \log(1 - x)$ and in accordance with a classical result \cite{KimuraandOhta1969} for $B=1$. For $x\to{}0$, we obtain
\begin{align}
\label{cTTF}
\bar{t^*}&=\frac{2\,B}{\sigma(e^{2\,B\,\sigma}-1)}\big((e^{2\,B\,\sigma}+1) \gamma-\textnormal{Ei}(2\,B\,\sigma)+\log(2\,B\,\sigma)\nonumber\\
&\hspace*{10pt}+e^{2\,B\,\sigma}(-\textnormal{Ei}(-2\,B\,\sigma)+\log(2\,B\,\sigma))\big),\hspace*{0.8cm}\sigma>0,\\
\bar{t^*}&=2\,B^2,\hspace*{0.8cm}\sigma=0,\nonumber
\end{align}
where $\gamma$ is Euler's constant and Ei denotes the exponential integral function \cite{abramowitzandstegun}.\par

\begin{figure*}[htbp]
\begin{center}
\includegraphics*[width=0.4\textwidth]{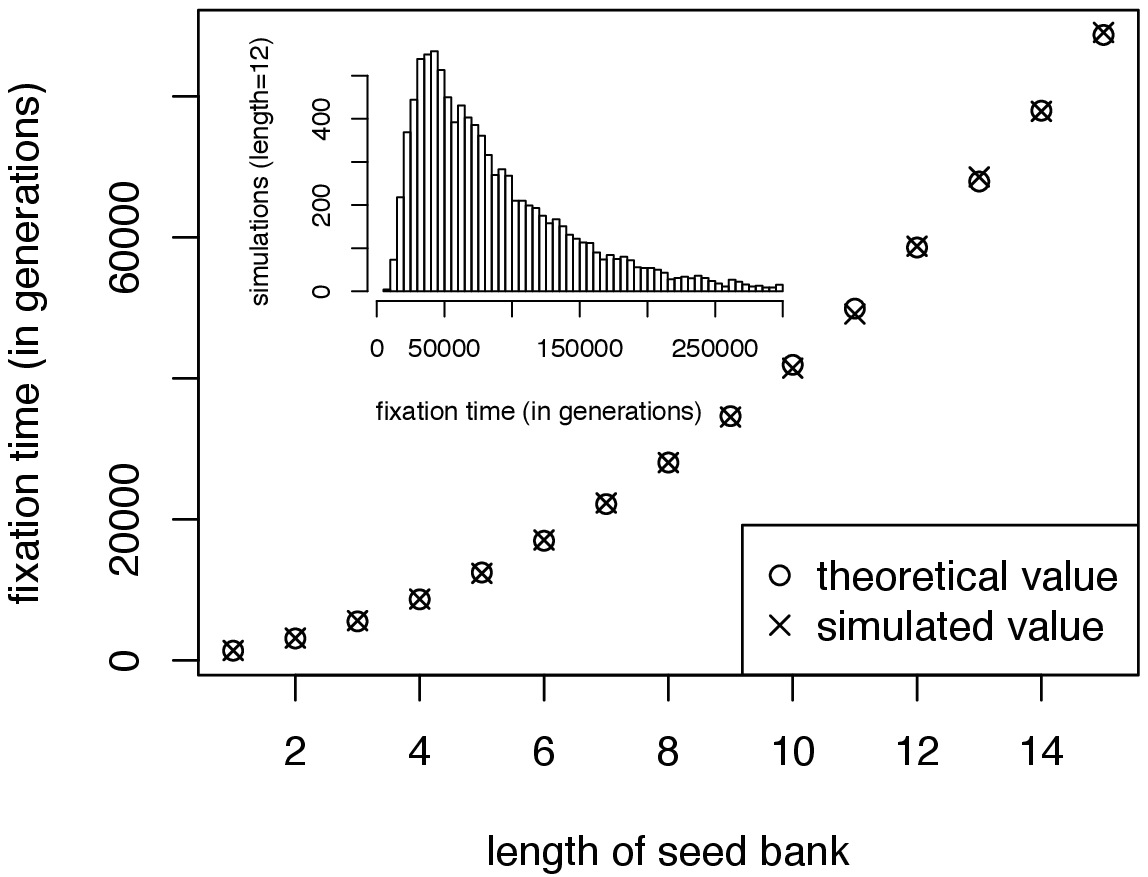}\hspace*{1cm}
\includegraphics*[width=0.4\textwidth]{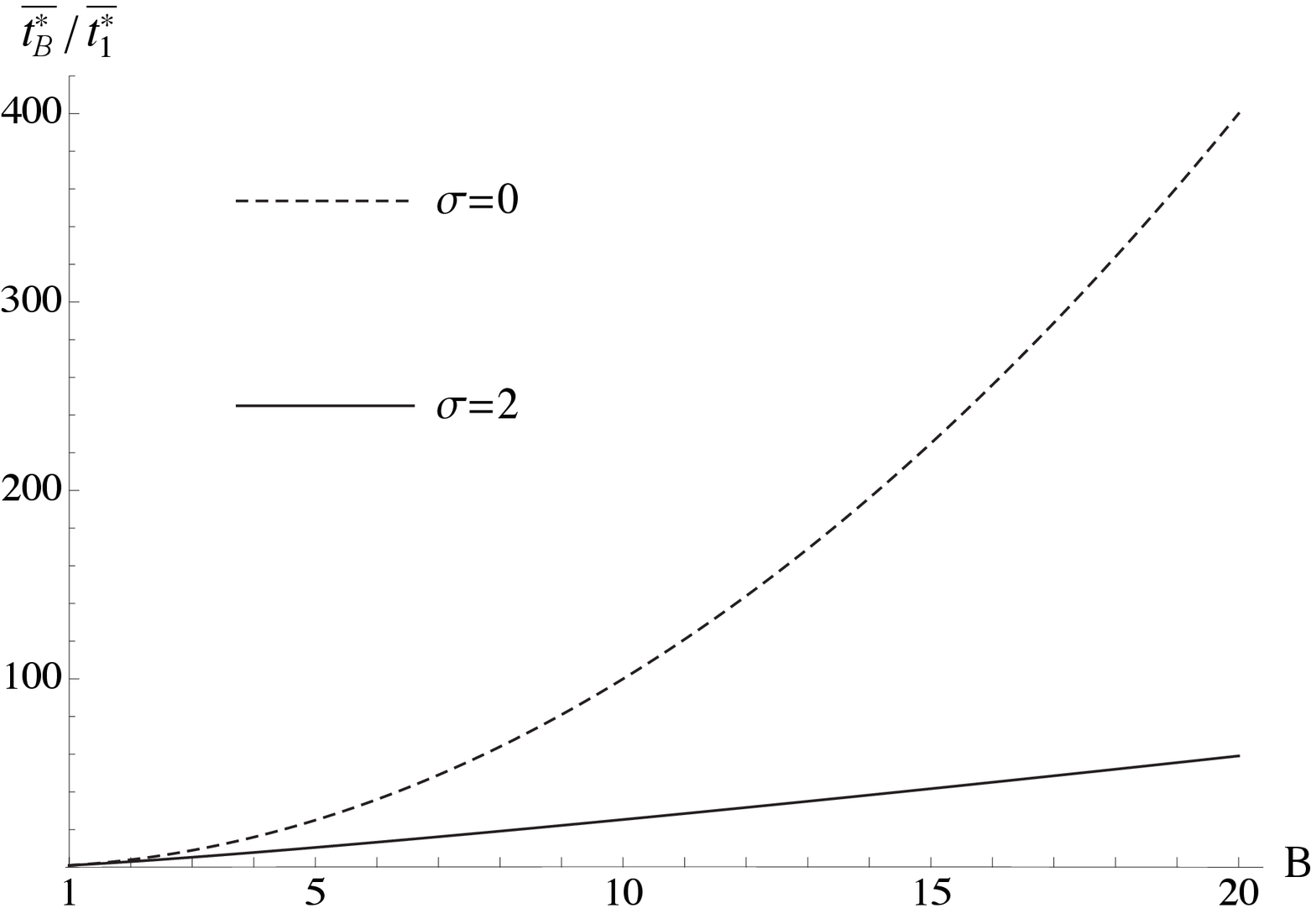}
\end{center}
\caption{(left) Simulation and theoretical prediction for the time to fixation of a seed bank model. The population size is 1000 and 50\% of the individuals are initially of genotype A. We simulated 10,000 runs for each mean value. The simulated distribution of the time to fixation is shown in the histogram at the upper left corner taking the data of the simulated seed bank of length $B=12$. (right) The ratios of the conditional fixation times with and without seedbank are plotted against the length of the seed bank $B$ for neutrality and selection by employing~(\ref{cTTF}). The additional index in the ratio is used to formally distinguish the cases with and without seed bank. }
\label{simSeedTE}
\end{figure*}

In Figure~\ref{simSeedTE} (left), we compare the time to absorption of the original discrete seed bank model by means of simulations with the theoretical result obtained from the diffusion approximation. For $b_A$ we use uniform distributions, where we vary the expected values between 1 and 8 corresponding to the length of the seed banks between 1 and 15. We choose an initial fraction of 0.5 for the type-A genotypes. The simulations show a good agreement between our analytical approximation and the numerical simulations. In Figure~\ref{simSeedTE} (right), we show the effect of the seed bank on the times to fixation conditional on fixation of the type-A genotype for neutrality and positive selection. 

\section{Discussion}

Within this study, we develop a forward in time Fisher-Wright model of a deterministically large seed bank with drift occurring in the above-ground population. The time that seeds can spend in the bank is bounded and finite, as assumed to be realistic for many plant or invertebrate species. We demonstrate that scaling time in the diffusion process by a factor $B^2$ generates the usual Fisher-Wright time scale of genetic drift with $B$ being defined as the average amount of time that seeds spend in the bank. The conditional time to fixation of a neutral allele is slowed down by a factor $B^2$ (Figure~\ref{simSeedTE} (right), dotted line) compared to the absence of seed bank. These results are consistent with the backward in time coalescent model from \citet{kaj2001}, and differs from the strong seed bank model of~\citet{blath2015a}. We evaluate the SFS based on our diffusion process and confirm agreement to the SFS obtained under discrete time Fisher-Wright simulations.\par
In the second part of the study, we introduce selection occurring at one of the two alleles, mimicking positive or negative selection. Two features of selection under seed banks are noticeable. First, selection is slower under longer seed banks (Figure~\ref{simSeedTE} (right), solid line) confirming previous intuitive expectations \cite{hairston1988}. Second, when computing the SFS with $B=2$ and without seed bank ($B=1$) under 
positive selection ($\sigma=2$) we reveal a stronger signal of selection for the seed bank by means of an amplified uptick of high-frequency derived variants. This effect becomes more prominent with longer seed banks and also holds for purifying selection, under which an increase in low-frequency derived variants is induced by the seed bank. We explain this counterintuitive results as follows: longer seed banks increase, on the one hand, the selection coefficient $\sigma$ generating a stronger signal at equilibrium (Figure~\ref{simSeedSfsSimVsTheo} (right)), and on the other hand, the time to reach this equilibrium state (Figure~\ref{simSeedTE} (right)). Our predictions are consistent with the inferred strengths of purifying selection in wild tomato species. Indeed, purifying selection at coding regions appears to be stronger in {\it S. peruvianum} than in its sister species {\it S. chilense} \cite{tellier2011} with {\it S. peruvianum} exhibiting a longer seed bank \cite{tellier2011a}.

\begin{acknowledgements}
{\it This research is supported in part by Deutsche Forschungsgemeinschaft grants TE 809/1 (AT) and STE 325/14 from the Priority Program 1590 (DZ).} 
\end{acknowledgements}

\begin{appendix}
\renewcommand\thesection{\appendixname\ \Alph{section}}

\section{Appendix: Moran model with deterministic seed bank}
\label{moranAppendix}
\renewcommand\thesection{\Alph{section}}
We briefly sketch the arguments that allow to handle
a Moran model with seed bank; the reasoning is completely 
parallel to the time-discrete case. In order
to keep this appendix short, we do not take into
account selection but focus on the neutral model.
\subsection{Model}
We start off with the individual based model. Let the
population size be $N$, $X_t$ the number of
genotype-A-plants, $\delta$ the death rate, and $b(s)$ 
the distribution of the 
ability for a seed at age $s$ to germinate; we require 
$\int_0^\infty b(s)\, ds=1$,  $B=\int_0^\infty s\,b(s)\, ds<\infty$, and $b(s)$ sufficiently smooth. Then, the rate for the transition $X_t\rightarrow X_t+1$ 
is given by 
\begin{eqnarray}
\delta\,N\, \hspace*{-2pt}\left(1-X_{t}/N\right)\hspace*{0pt}\int_0^{\infty} \hspace*{-2pt}
b(\tau) \, X_{t-s}/N ds,
\end{eqnarray}
while that for a decrease of $X_t$ by $1$ reads
\begin{eqnarray}
\delta\,N\,\hspace*{-2pt}\left(X_{t}/N\right)\hspace*{-2pt}\left(\hspace*{0pt}1\hspace*{-1pt}-\hspace*{-3pt}
 \int_0^{\infty}\hspace*{-2pt}b(\tau) X_{t-s}/N ds\hspace*{0pt}\right).
\end{eqnarray}
\begin{eqnarray}
&&P(X_{t+\Delta t}
 = X_t+1|X_\tau\mbox{ for } \tau\leq t)\\
 &=&\Delta  t\,\delta\,N\, \hspace*{-2pt}\left(1-X_{t}/N\right)\hspace*{0pt}\int_0^{\infty} \hspace*{-2pt}
b(\tau) \, X_{t-s}/N ds+{\cal O}(\Delta t),\nonumber\\
&&P(X_{t+\Delta t}  = X_t-1|X_\tau\mbox{ for } \tau\leq t)\\
& &=\Delta  t\,\delta\,N\,\hspace*{-2pt}\left(X_{t}/N\right)\hspace*{-2pt}\left(\hspace*{0pt}1\hspace*{-1pt}-\hspace*{-3pt}
 \int_0^{\infty}\hspace*{-2pt}b(\tau) X_{t-s}/N ds\hspace*{0pt}\right)+{\cal O}(\Delta t).\nonumber
\end{eqnarray}
Note that the delay process requires the knowledge 
of the complete history $\{X_s\}_{s<t}$. The usual continuous 
limit for $u_t=X_t/N$ yields (with $\eps = 1/N$)
\begin{eqnarray*}
du_t &=&
\delta\,\, \left(\,\,\int_0^{\infty} 
b(s) \, u_{t-s}\, ds 
-\,\,u_{t}\right)\, ds \\
&&+ 
\left\{\eps \delta\,
\int_{0}^\infty b(s) 
(u_t 
+ u_{t-s}
 - 2 u_t  \, u_{t-s})\,ds\right\}^{1/2}
 d W_t.
\end{eqnarray*}
If we rescale time in the usual way, 
$\tau=\eps t$, and define  $v_\tau = u_{\tau/\eps}$, we obtain
\begin{eqnarray}
&&dv_\tau = \eps^{-1}\,\,\delta 
\left(\eps^{-1}\,
 \int_0^{\infty}  b(s/\eps)(v_{\tau-s}-v_\tau)\,ds\right)\, d\tau \label{scaledModel}\\
 &&
+ \left(\eps^{-1}\,
\delta\,\int_0^{\infty} b(s/\eps)
\big(v_\tau+v_{\tau-s}
-2\,v_\tau\,v_{\tau-s}\big)\,ds\,\right)^{1/2}\, dW_\tau.\nonumber
\end{eqnarray}
The aim here is to find heuristic arguments 
indicating that $v_\tau$ 
approximates for $\eps\rightarrow 0$ the solution of 
a Moran diffusion 
process with
rescaled time, paralleling equation~(\ref{moranDiffProcess}). \par\medskip
Note that, 
in some sense, the terms in this time-continuous model 
are better to interpret than the parallel terms in the Fisher-Wright model: both terms within the brackets are moving averages, and clearly 
\begin{eqnarray}
&&\lim_{\eps\rightarrow 0}\left(\eps^{-1}\,
\delta\,\,\int_0^{\infty}b(s/\eps)
\big(u_\tau+u_{\tau-s}
-2\,u_\tau\,u_{\tau-s}\big)\,ds\,\right)\nonumber\\
&&=  
2\,\delta\,u_\tau(1-u_\tau)\label{scaledNoise}
\end{eqnarray}
for a function $u_\tau$ that is reasonably smooth. For the drift term, we find similarly 
\[
\lim_{\eps\rightarrow 0}\left(\eps^{-1}\,
\int_0^{\infty} b(s/\eps)\,(u_{\tau-s}-u_\tau)\,ds\right)\rightarrow u_\tau-u_\tau = 0.
\]
However, in eqn.~(\ref{scaledModel}), this bracket is divided by $\eps$, and hence does
not vanish for $\eps\rightarrow 0$. If we take a closer look, we
find that a deviation of $u_\tau$ from the moving average 
(the state of the seed bank) is punished. That is, the state of 
living plants can change only slower in comparison with a model 
without seed bank, and therefore for $\eps\rightarrow 0$ we expect 
a diffusion model at a slower time scale. \par\medskip
{\it Remark:} At this point we may use a formal argument that parallels that for approximations of SDDE with a small delay by an SDE in~\cite{Guillouzic1999}: For a smooth function  $\psi$, we may write 
\begin{eqnarray*}
&&\eps^{-1}\int_0^{\infty} b(s/\eps)\,\psi(-s)\,ds \\
&=&\eps^{-1} \int_0^{\infty} b(s/\eps)\,(\psi(0)-s\psi'(0)+{\cal O}(s^2))\,ds \\
&=& \psi(0) - \eps\, \psi'(0)\, B+{\cal O}(\eps^2)
\end{eqnarray*}
and hence, in a very formal sense, we may refine the 
considerations above for the drift term, 
\begin{eqnarray}
\lim_{\eps\rightarrow 0}\left(\eps^{-2}\,
\int_0^{\infty} b(s/\eps)\,(u_{\tau-s}-u_\tau)\,ds\right)\, dt\rightarrow -B\,du_\tau 
.\qquad
\end{eqnarray}
Combining this result with equations~(\ref{scaledModel}), (\ref{scaledNoise}) yields
$ dv_\tau (1+\delta B) = \left(2\delta v_\tau(1-v_\tau)\right)^{1/2}\, dW_\tau$ and hence
\begin{eqnarray}
dv_\tau = \frac{(2\delta v_\tau(1-v_\tau)^{1/2}}{1+B\,\delta}\, dW_\tau
.
\end{eqnarray}
This argument is nice and short but this formal that it requires a less formal support. We indicate this supporting computation in the next section. 

\subsection{Scaling $\eps\rightarrow 0$}
In order to use the arguments developed in the main part of the article, we
discetize the stochastic differential-delay equation by the
Euler-Maruyama formula, and find  
\begin{eqnarray*}
&&v_{\tau+\Delta \tau} 
= v_\tau -
\eps^{-1}\,\delta\,\Delta \tau \left(v_\tau -  \sum_{i=1}^\infty\,
v_{\tau-i\Delta \tau} \varphi_i^{(\Delta \tau)}\right)\\ 
&&+\left(
\delta\, \sum_{i=1}^\infty \varphi_i^{(\Delta \tau)}
\big(v_\tau+v^\eps_{{t-i\Delta \tau}}
-2\,v_\tau\,v^\eps_{{\tau-i\Delta \tau}}\big)\,\right)^{1/2}
\sqrt{\Delta \tau}\, \eta_\tau,
\end{eqnarray*}
where $\eta_\tau$ are i.i.d. $N(0,1)$ distributed, and 
the weights $\varphi_i^{(\Delta t)}$ are chosen as 
\[
\varphi_i^{(\Delta \tau)} = b(i\, \Delta \tau/\eps)
(\Delta \tau/\eps)+{\cal O}(\Delta \tau^2/\eps),
\]
such that $\sum_{i=1}^\infty\, \varphi_i^{(\Delta \tau)} = 1$.
  If we now define 
  \begin{eqnarray*}
  \beta &=& \left(
\delta\, \sum_{i=1}^\infty \varphi_i^{(\Delta \tau)}
\big(v_\tau+v^\eps_{{t-i\Delta \tau}}
-2\,v_\tau\,v^\eps_{{\tau-i\Delta \tau}}\big)\,\right)^{1/2},\\
\psi(x) &=& 1-z+\delta\Delta \tau \eps^{-1}\,\,\left(z-\sum_{i=1}^\infty \varphi_i^{(\Delta t)} z^{i+1}\right),
  \end{eqnarray*}
  we may rewrite the discretized equation for $v_\tau$ as 
\[
\psi(L) v_{\tau+\Delta\tau} = \beta \sqrt{\Delta \tau}\, \eta_\tau,
\]
  where $L v_\tau = v_{\tau-\Delta\tau}$. 
We are now in the position to apply 
the computations about the quasi-stationary state of the seedbank (neglecting the time-dependency of $\beta$). 
As 
\begin{eqnarray*}
 -\psi'(1) & = & 
1 -\delta\Delta \tau/\eps + \delta\,\sum_{i=1}^\infty \varphi_i^{(\Delta t)} (i+1) \Delta \tau/\eps \\
&=&
1 -\delta\Delta \tau/\eps 
+ \delta\,\sum_{i=1}^\infty 
 b(i\, \Delta \tau/\eps) (i \Delta \tau/\eps) 
(\Delta \tau/\eps) \\
&&\hspace*{13pt}+  \Delta \tau/\eps \,\delta\,\sum_{i=1}^\infty 
\left( b(i\, \Delta \tau/\eps)
(\Delta \tau/\eps)+{\cal O}(\Delta \tau^2/\eps)\right),
\end{eqnarray*}
we have
\[
1+\delta\int_0^\infty b(s)\, s\, ds
= 1+\delta B\quad\mbox{for \,} \Delta\tau/\eps\rightarrow 0,
\]
and conclude that approximately 
\[
v_{\tau+\Delta\tau} = v_\tau + 
\frac{\beta \sqrt{\Delta \tau}}{1+\delta\, B}\, \eta_\tau.
\] 
Hence, for $\eps\rightarrow 0$ we expect 
(according to these heuristic arguments) that $v^\eps_\tau$ 
satisfies the rescale diffusion equation
\begin{eqnarray*}
 dv_\tau = 
 \frac {(2\,\delta v_\tau(1-v_\tau))^{1/2}} {1+\delta\, B}\, dW_\tau.
\end{eqnarray*}
If we define $G=1/\delta$, the average inter-generation time
of living plants, this equation becomes even closer to that derived for the Fisher-Wright case,
\begin{eqnarray}
 dv_\tau = 
 \frac {(2\,\delta v_\tau(1-v_\tau))^{1/2}} {(1+B/G)}\, dW_\tau
\end{eqnarray}
as it becomes clear that the correction factor 
$1+B/G$ measures the average time a seed rests in the soil
in terms of generations. 
\end{appendix}


\bibliographystyle{apsrmp4-1}       

\bibliography{popGen}


\end{document}